\documentclass[letterpaper,10pt,journal,twocolumn]{IEEEtran}

\IEEEoverridecommandlockouts

\usepackage{color}
\usepackage[active]{srcltx}

\usepackage{epsfig}
\usepackage{amsmath}
\usepackage[flalign]{mathtools}
\usepackage{amssymb}
\usepackage{psfrag}
\usepackage{graphicx}
\usepackage{txfonts}
\usepackage{tikz}
\usetikzlibrary{arrows}
\usepackage{ marvosym }
\usepackage{epstopdf}

\usepackage{color}
\usepackage{tikz}
\usetikzlibrary{arrows}

\graphicspath{{../../figure/}}

\newtheorem{dfn}{Definition}
\newtheorem{thm}{Theorem}[section]

\newtheorem{obs}{Observation}
\newtheorem{prop}[thm]{Proposition}
\newtheorem{example}[thm]{Example}

\ifCLASSOPTIONcompsoc
  \usepackage[nocompress]{cite}
\else
  \usepackage{cite}
\fi
\ifCLASSINFOpdf
\else
\fi

\usepackage[]{footmisc}

\begin{document}

\newcommand{\ComplexField}{\mathbf{C}}
\newcommand{\NaturalNumbers}{\mathbf{N}}

\newcommand{\eps}{\varepsilon}
\newcommand{\Eps}{\mathcal{E}}

\newcommand{\w}{\omega}
\newcommand{\C}{\mathbf{C}}
\newcommand{\Z}{\mathbb{Z}}
\newcommand{\I}{\mathcal{I}}
\newcommand{\N}{\mathbb{N}}
\newcommand{\R}{\mathbb{R}}

\newcommand{\tfspan}{\mathrm{tfspan}}
\newcommand{\ctfspan}{\mathrm{ctfspan}}

\newcommand{\ZF}{~{}^{0}\mathcal{F}}

\newcommand{\node}{y}
\newcommand{\nodenumber}{n}
\newcommand{\nodeindex}{j}
\newcommand{\nodeindexalt}{i}
\newcommand{\nodeindexaltb}{k}
\newcommand{\firstnodeindex}{1}
\newcommand{\secondnodeindex}{2}
\newcommand{\thirdnodeindex}{3}
\newcommand{\lastnodeindex}{\nodenumber}
\newcommand{\collider}{c}
\newcommand{\collidernumber}{n_c}
\newcommand{\ancestornumber}{n_a}

\newcommand{\path}{\pi}
\newcommand{\pathfirstnode}{0}
\newcommand{\pathsecondnode}{1}
\newcommand{\pathsecondlastnode}{l-1}
\newcommand{\pathlastnode}{l}
\newcommand{\pathlength}{\ell}
\newcommand{\pathnodeindex}{p}
\newcommand{\pathnodeindexalt}{q}
\newcommand{\pathapex}{(a)}
\newcommand{\pathapexalt}{(b)}

\newcommand{\ov}{\overline}

\newcommand{\parent}[2]{pa_{#1}\left(#2\right)}
\newcommand{\child}[2]{ch_{#1}\left(#2\right)}
\newcommand{\ancestor}[2]{an_{#1}\left(#2\right)}
\newcommand{\descendant}[2]{de_{#1}\left(#2\right)}

\newcommand{\cov}{R}
\newcommand{\timelag}{\tau}
\newcommand{\E}{\mathbf{E}}
\newcommand{\Ztrans}{\mathcal{Z}}

\newcommand{\graph}{G}
\newcommand{\vertexset}{V}
\newcommand{\edgeset}{E}

\newcommand{\subsetvertex}{J}
\newcommand{\subsetvertexalt}{I}
\newcommand{\subsetvertexsep}{Z}

\newcommand{\imap}{\mathcal{I}}
\newcommand{\granger}{\mathcal{G}}
\newcommand{\csep}{\mathcal{C}}

\newcommand{\colliderset}{C}

\newcommand{\PSD}{\Phi}

\newcommand{\TheoremMatSalMarkovBlanketnoncausal}{Theorem~27 in \cite{MatSal12}}
\newcommand{\LemmaMatSalWienerUncorrelated}{Lemma~26 in \cite{MatSal12}}

\newcommand{\childrenset}{\mathcal{C}}
\newcommand{\parentset}{\mathcal{P}}
\newcommand{\coparentset}{\mathcal{K}}
\newcommand{\tf}{H}

\newcommand{\lemmarestrictiontoancestors}{Lemma~6 in \cite{MatSal14}}
\newcommand{\lemmaenlargeseparatedsets}{Lemma~7 in \cite{MatSal14}} 

\newcommand{\wiener}{W}
\newcommand{\Wiener}[3][]{
	\wiener_{
		#2
		\ifx&#1&
		\else
		,[#1]
		\fi
		|#3}
	}

\newcommand{\LDIMGraph}{Direct Feedthrough Graphical Representation }
\newcommand{\LDIMgraph}{direct feedthrough graphical representation }

\title{
    Granger-faithfulness and link orientation in network reconstruction$^{*}$ \thanks{$^{*}$Under review at IEEE Transactions on Control of Network Systems (TCNS)}
}

\author{
Mihaela Dimovska \& Donatello Materassi ({\tt\small dimov003@umn.edu \& mater013@umn.edu})\\
		Department of Electrical and Computer Science,\\
		University of Minnesota,
 		200 Union St SE, 55455, Minneapolis (MN) 
}

\maketitle

\begin{abstract}

Networked dynamic systems are often abstracted as directed graphs, where the observed system processes form the vertex set and directed edges are used to represent non-zero transfer functions.  
Recovering the exact underlying graph structure of such a networked dynamic system, given only observational data, is a challenging task. 
Under relatively mild well-posedness assumptions on the network dynamics, there are state-of-the-art methods which can guarantee the absence of false positives. 
However, in this article we prove that under the same well-posedness assumptions, there are instances of networks for which any method is susceptible to inferring false negative edges or false positive edges.
Borrowing a terminology from the theory of graphical models, we say those systems are unfaithful to their networks. 
We formalize a variant of faithfulness for dynamic systems, called Granger-faithfulness, and for a large class of dynamic networks,
we show that Granger-unfaithful systems constitute a Lebesgue zero-measure set. 
For the same class of networks, under the Granger-faithfulness assumption, we provide an algorithm that reconstructs the network topology with guarantees for no false positive and no false negative edges in its output.
We augment the topology reconstruction algorithm with orientation rules for some of the inferred edges, and we prove the rules are consistent under the Granger-faithfulness assumption.
\end{abstract}

\section{Introduction}

Learning the underlying connectivity of a network of dynamic systems only from observational data is of interest in a variety of application domains, such as 
economics \cite{naylor2007topology}, 
biology \cite{del2015biomolecular}, neuroscience \cite{kaminski2001evaluating}, 
or artificial intelligence \cite{gong2017causal}. 
In most applications, the system under study can be abstracted via a causal diagram. 
The causal diagram is a directed graph where the vertices represent the measured output processes and the presence of a link represents a form of causal influence from one process to another.
One of the prevalent methods for inferring the existence of a direct causal influence between two observed processes is Granger causality \cite{granger1969}.
Inspired by the work of Norbert Wiener on the estimation of time series \cite{wiener1949}, Clive Granger proposed an operational notion of causality which can be effectively implemented via statistical tests.
Namely, Granger causality tests if the past observations of a variable $y_{i}$ contribute significantly to the prediction of the variable $y_{j}$, given the past observations of all the other output variables in the system except $y_{i}$ \cite{granger1969}.
In such a case, it is said that ``$y_{i}$ Granger-causes $y_{j}$'' and a directed edge from $y_{i}$ to $y_{j}$ is inferred in the causal diagram.
For a large class of networks, namely in the linear case and when all the transfer functions of the system are strictly causal,
 Granger causality is proven to be consistent \cite{sims1980macroeconomics, materassi2012problem, gong2017causal, GonWar08}.
Similarly, the Dynamical Structure Function (DSF) framework \cite{GonWar08}, when formulated for the reconstruction of dynamic networks from observational time series (blind reconstruction), provides guarantees of a consistent reconstruction for systems with strictly causal dynamics  \cite{hayden2016network}.
However, it is well-known that Granger causality and similar methods can lead to the inference of false positive/negative links in the presence of direct feedthroughs or confounders \cite{barnett2015granger, eichler2007granger}, limiting their applicability to wider scenarios.

\indent On the other hand, methods that consider networks with direct feedthrough components face several substantial challenges, one of them being the challenge of providing consistency guarantees. 
Many network reconstruction methods that take into account the presence of direct feedthroughs are susceptible to inferring both false positive and false negative edges in the reconstructed network \cite{Quinn2015directed, schiatti2015extended, runge2012escaping}. 
However, there are a few methods, designed for static \cite{spirtes2000causation} or dynamic systems \cite{materassi2012problem, dimovska2017granger} that provide guarantees to obtain no false positives under the assumption that there are no algebraic loops in the networks. 
This article illustrates how the inference of false negatives is more delicate. 
As a first contribution of this article, we show that, even in the absence of algebraic loops, there are instances of networks for which any method is susceptible to inferring false negative or false positive edges. 
In those instances, borrowing a terminology from the theory of graphical models, we say that the dynamic system is not ``faithful'' to its causal network \cite{spirtes2000causation}.
It can definitely be argued that unfaithfulness can have practical implications, especially for small data sizes \cite{uhler2013geometry}. 
However, for a consistency analysis, which is the main focus of this article, we claim that unfaithfulness is of limited concern. 
To that end, we consider a large class of parameterized dynamic systems for which we define a notion of faithfulness, called Granger-faithfulness. 
For this class, we prove that 
the set of parameters that lead to Granger-unfaithful systems is a Lebesgue zero-measure set.
Therefore, assuming that a system is Granger-faithful can be seen as a technical assumption aiming at removing a zero-measure set of parameters associated with pathological situations.

For the same class of parameterized networks, we modify the topology reconstruction algorithm in \cite{dimovska2020control} to obtain a consistent algorithm that under Granger-faithfulness has guarantees for no false positive or negative edges in its output.

As an additional contribution, we look at the issue of orienting edges in the inferred topology and we show that Granger-faithfulness is a useful concept for orienting edges with direct feedthroughs.
When the network has strictly causal dynamics, inferring that $y_{i}$ Granger-causes $y_{j}$ naturally implies that the inferred link between $y_{i}$ and $y_{j}$ has a causal direction from $y_{i}$ to $y_{j}$. 
However, when direct feedthroughs are present, such ``time-given'' information is not available and atemporal methods for orientation inference need to be used.
In this article, we demonstrate that atemporal notions of causality from the area of graphical models can be used to infer the causal direction of some influences associated with non zero direct feedthroughs. 
We augment the topology reconstruction algorithm in \cite{dimovska2020control} with orientation rules and we prove that the rules are consistent under the Granger-faithfulness assumption.

\indent The article is structured as follows: in Section~\ref{sec: background}  we specify the model of networks we consider in this work as well as the necessary background information; 
In Section~\ref{sec: counterexample false negatives} we illustrate an instance of a well-posed, Granger-unfaithful network that cannot be reconstructed exactly by any method;
In Section~\ref{sec: GEMD and Granger-faithfulness} $A$ we define the notion of Granger-faithfulness and we present a variation of the method in \cite{dimovska2020control}.
In Section~\ref{sec: GEMD and Granger-faithfulness} $B$ we show that unfaithfulness is of limited concern as the systems that are Granger-unfaithful constitute a zero measure set. 
In Section~\ref{sec: orientations} we augment the algorithm provided in Section~\ref{sec: GEMD and Granger-faithfulness} $A$ with orientation rules that we also show to be consistent. 

\section{Background and Problem Statement}\label{sec: background}


In this work we develop our results for a class of dynamic stochastic networks called Linear Dynamic Influence Models (LDIMs), which have been extensively studied \cite{materassi2019signal, dimovska2017granger, dimovska2020control}. 
These networks can be considered as a special case of Dynamic Structure Function \cite{GonWar08} with unknown (not measured) forcing signals that are modeled as mutually independent stochastic processes.
We first recall the LDIM definition. 
\begin{dfn}[Linear Dynamic Influence Model]
	A Linear Dynamic Influence Model $\mathcal{G}$ is a pair $(H(z),e)$ where
	\begin{itemize}
		\item	$e=(e_1, ..., e_N)^{T}$ is a vector of $N$ processes $e_{1}$, ... $e_{N}$, such that 
		$\Phi_{e}(z)$, the Power Spectral Density (PSD) of $e$,
		is real-rational and diagonal, namely $\Phi_{e_{i}e_{j}}=0$ for $i\neq j$.
		\item	$H(z)$ is an $N\times N$ real-rational transfer matrix.    $H(z)$ is termed as the ``dynamics'' of the LDIM. We denote the $ji$-th entry of $H(z)$ by $H_{ji}(z)$.
	\end{itemize}
	The output processes $\{y_j\}_{j=1}^{N}$ of the LDIM are defined as
	$y_j=e_j+\sum_{i=1}^{N}H_{ji}(z)y_i$,
	or in a matrix notation
	$
		y=e+H(z)y
	$,
	where $y=(y_{1}, ..., y_{N})^{T}$.
	Note that we could further write $e=F(z)u$, where $F(z)=diag(F_{1}(z),...,F_{N}(z))$ is a diagonal matrix of minimum phase biproper transfer functions and $u=(u_{1},...,u_{N})$ is a vector of $N$ scalar processes such that $\Phi_{u}(z)$ is the identity matrix. 
\end{dfn}
Further assume the following:
\begin{itemize}
 \item Every entry of $H(z)$ is causal. 
 \item  For every set of indices $I \subseteq \{1,...,n\}$, 
 $(I-H_{II}(z))^{-1}$ exists and is causal. 
 \item $\Phi_{e_ie_i}(e^{i\omega}) > 0$ for any $\omega \in \left[-\pi, \pi \right]$ and for any $i = 1, ..., n$. 
 \item There are no algebraic loops in the dynamics $H(z)$. Thus, in this work we develop our results for the so called \textit{recursive} LDIMs, i.e. LDIMs that have at least one strictly causal transfer function in each feedback loop \cite{dimovska2017granger}.
\end{itemize}

We can associate a directed graph representation to a LDIM, called the causal graph of the LDIM, where the connectivity of the corresponding directed graph matches the sparsity of the dynamics matrix, $H(z)$. 

\begin{dfn}[Causal Graph of a LDIM]
Let $\mathcal{G}$ be a LDIM. Let $G=(V,E)$ be a directed graph defined as follows:
the vertex set $V=\{y_{1},...,y_{n}\}$ is the set of the output processes of the LDIM
and the ordered pair $(y_{i}, y_{j}) \in E$ if and only if $H_{ji} \neq 0$.
We refer to $G$ as the causal graph of the LDIM.
\end{dfn}

Utilizing the notion of a causal graph of a LDIM, we can formulate the problem we are interested in solving as a graph learning problem. 

\section*{Problem Statement: Causal Graph Reconstruction}

\textbf{Given observational data from the output processes $y$ of a recursive LDIM, learn the causal graph of the LDIM, with orientations.}

\subsection{Graphs associated with LDIMs}

To learn the causal graph of a LDIM we introduce additional notions from graph theory and graphical models. These notions will help us consider the dynamics of a LDIM in a more refined way, by allowing us to distinctly represent strictly causal dependencies and dependencies with direct feedthroughs. 
To this end, we make use of both directed and undirected graphs. 
For directed graphs, we use the standard notions of children, parents, ancestors and descendants (see \cite{materassi2019signal}). 
For a directed graph $G$, we further use the notions of coparents and colliders, which are defined below.

 \begin{dfn}[Coparents, collider]
 Let $G=(V,E)$ be a directed graph. 
 Two nodes $y_{i}$ and $y_{j}$ are coparents 
 if and only if $y_{i}$ and $y_{j}$ share at least one common child $y_{k}$. 
 We say that $y_{k}$ is a collider between $y_{i}$ and $y_{j}$.
 \end{dfn}
 
Note that the causal graph of a LDIM is a standard directed graph that potentially contains directed cycles.
To suggestively represent links with strictly casual dynamics and links with direct feedthrough dynamics, we further make use of so called directed multi-arrowed graphs, which are extensions of directed graphs \cite{dimovska2020control, shafie2015multigraph}. 

\begin{dfn}[Graphical representation of a LDIM \cite{dimovska2020control}] 
	Let
	$\mathcal{G}=(H(z),e)$
	be a LDIM with output processes
	$y_{1},...,y_{n}$.
	Let
	$V:=\{y_{1},..., y_{n}\}$
	and let
	$E_{1}$
	and 
	$E_{2}$
	be two disjoint subsets of
	$V \times V$ such that
    \begin{itemize}
		\item[(a)] $(y_{i}, y_{j}) \notin E_{1} \cup E_{2}$ implies $H_{ji}=0$
		\item[(b)] $(y_{i}, y_{j}) \notin E_{1}$ implies $H_{ji}(z)$ is strictly causal. 
	\end{itemize}
	We say that the multi-arrowed graph
	$G=(V,E_{1},E_{2})$
	is a graphical representation of the LDIM.
	Furthermore, if the implications (a) and (b) hold also in the opposite direction, we say that  
	$G=(V,E_{1},E_{2})$
	is a perfect graphical representation of the LDIM.
\end{dfn}

Note that all the standard notions of parents, children, ancestors and descendants of a node in a directed graph extend to a node in a graphical representation of a LDIM. 
Further, observe that there is a subtle but important distinction between a graphical representation of a LDIM and the perfect graphical representation of a LDIM. 
Namely, given a \textit{recursive} LDIM, the complete graph with only single-headed arrows is always a graphical representation of the LDIM. 
However, the \textit{perfect} graphical representation of a recursive LDIM must be such that there is at least one double-headed arrow in every directed cycle. 
In this work, we aim to provide a method for inferring the perfect graphical representation of a recursive LDIM, as then it is immediate to obtain the causal graph of the LDIM, as the following observation states. 
\begin{obs}
Let $(H(z),e)$ be a LDIM, with perfect graphical representation $G=(V,E_{1}, E_{2})$.
Then its causal graph is $G^{c}=(V, \{E_{1}\cup E_{2}\})$.
\end{obs}

Next, we consider a subgraph of a graphical representation of a LDIM that is associated only with the single-headed edges of a graphical representation. 

\begin{dfn}[Graph of instantaneous propagations]\label{dfn:Graph instantaneous propagations}
	Consider a recursive LDIM $(H(z),e)$ with outputs
	$y_1, ..., y_n$.
	Let $G$ be a graphical representation of the LDIM.
	The graph of instantaneous propagations of $G$ is the graph $G^{\text{\Lightning}}$
	where there is an edge from $y_{i}$ to $y_{j}$
	if and only if there is a single-headed link from $y_i$ to $y_j$ in $G$. 
\end{dfn}

We also consider the underlying undirected graph of a the causal graph of a LDIM. 
\begin{dfn}[Skeleton (Topology) ~\cite{Pea88}]
	Given a directed graph $\graph=(\vertexset,\edgeset)$, we define its ``skeleton'' or ``topology'' as the undirected graph $\overline{G} = (\vertexset,\ov \edgeset)$ obtained by removing the orientation of its edges. 
\end{dfn}

\begin{figure}[h!]
	\centering
	\bgroup
	\begin{tabular}{ccc}
	\includegraphics[width=0.35\columnwidth]{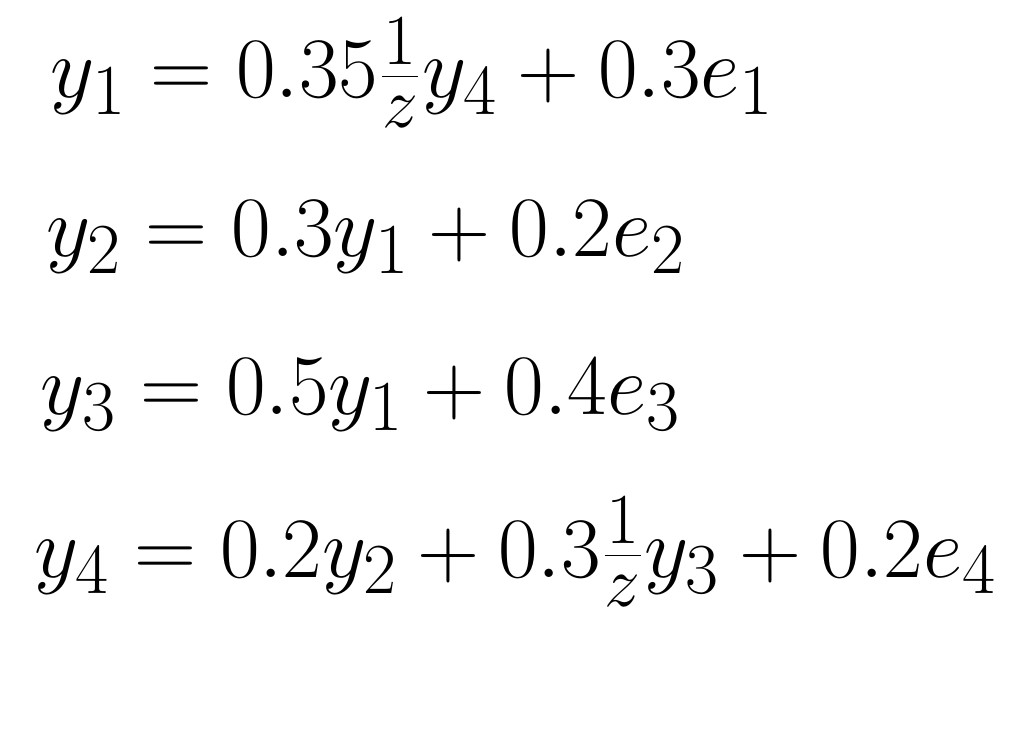} &
	\includegraphics[width=0.24\columnwidth]{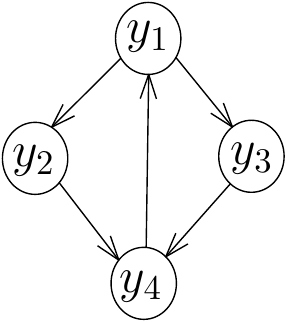} &
	\includegraphics[width=0.24\columnwidth]{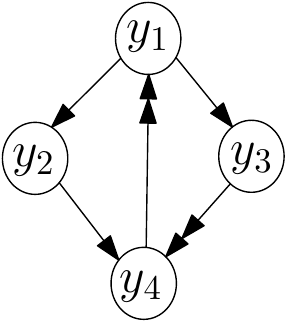}
	\\
	& & \\
	(a) & (b) &(c) \\
	& & \\
	 \includegraphics[width=0.24\columnwidth]{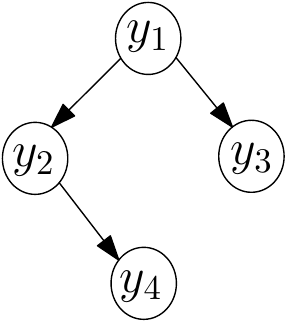} &  &
	 \includegraphics[width=0.24\columnwidth]{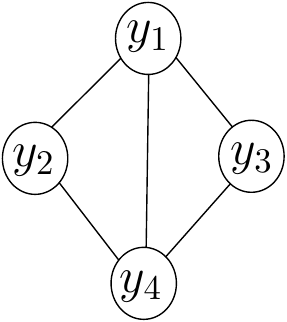} \\
	  \\ 
	(d) &  & (e)
	\end{tabular}
	\egroup
	\caption{ (a) Dynamics of the LDIM; (b) Causal graph (c) Perfect graphical representation (d) Graph of instantaneous propagations (e) Topology (skeleton).
	\label{fig:graph notions to LDIM} }
\end{figure}

 The graph notions associated with a LDIM (causal graph, perfect graphical representation, graph of instantaneous propagations, topology) are further illustrated in Figure~\ref{fig:graph notions to LDIM}. 

Next, we present some systems-theoretic tools formulated as Wiener filter operators, which we subsequently use to infer the graph notions associated with LDIMs. 

\subsection{Projection operators on LDIMs}

A mathematical framework describing properties and notions related to LDIMs is developed in \cite{dimovska2020control}.
Here we recall some important notions from that framework that will enable us to define projections on LDIM processes.
We start by recalling the notions of transfer function span ($tfspan$) and causal transfer function span ($ctfspan$). 

\begin{dfn}[(Causal) Transfer Function Spans]
Given a set of rationally related random processes $E$, the transfer function span of $E$ is:
\begin{align*}
		tfspan(E)&:=
			\left\{y=\sum_{k}H_{k}(z)e_{k}~\middle\vert~ e_{k}\in E\,, \right\} 
\end{align*}
 where $ H_{k}(z) $ are not necessarily causal transfer functions. 
 
Similarly, the causal transfer function span of $E$ is:
\begin{align*}
		ctfspan(E)&:=
			\left\{y=\sum_{k}H_{k}(z)e_{k}~\middle\vert~ e_{k}\in E\,, \right\} 
\end{align*}
 where $ H_{k}(z) $ are causal transfer functions.  
\end{dfn}

The (causal) transfer function spaces are shown to be pre-Hilbert spaces  \cite{materassi2012problem} when equipped with the inner product
\[
\left\langle y_i, y_j\right\rangle = \frac{1}{2\pi}\int_{-\pi}^{\pi}{\Phi_{y_iy_j}(e^{i\omega})d\omega} \,,
\]
which induces a norm in the usual way. 
The pre-Hilbert space structure enables us to use the tools of non-causal and causal Wiener filters and consider projections of rationally related processes onto $tfspan$/$ctfspan$ spaces. 
In fact, given a set of rationally related processes $y = \left(y_1,...,y_n\right)^{T}$, we can consider the standard least-squares problem to define the projection $\hat v$ on a space $Y$ as 
\[
\hat v= \operatorname*{arg\,min}_{q \in Y } \left\|v-q\right\|^2 \,,
\] 
where $ Y = tfspan\left(y_1,...,y_n \right)$ or $Y= ctfspan\left(y_1,...,y_n \right)$. 
A fundamental result states that $\hat v$ exists and is unique \cite{luenberger1997optimization}.  
When $Y=tfspan\left(y_1,...,y_n \right)$ we denote the solution $\hat v=W^{nc}(z)y$ and we refer to $W^{nc}(z)$ as the non-casual Wiener filter. 
Similarly when $Y=ctfspan\left(y_1,...,y_n \right)$ we denote the solution $\hat v=W^{c}(z)y$ and we refer to it as the casual Wiener filter. 
The expression of the (causal) Wiener filter is unique when $\Phi_{yy}(e^{i\omega})$ is positive definite for $\omega \in [-\pi, \pi]$ \cite{materassi2012problem}. 
For formal statements and detailed derivations of the non-causal/causal Wiener filter, refer to \cite{materassi2012problem}. 

These concepts lead to the notions of orthogonality and ``Wiener separation'' \cite{materassi2019signal}. 

\begin{dfn}[Orthogonality to $(c)tfspans$]

Let $v$ and $y_1,..., y_n$ be $n+1$ rationally related processes. 
We say that $v$ is orthogonal to $(c)tfspan(y_1,..., y_n)$, 
denoted as $v \perp (c)tfspan(y_1,..., y_n)$, 
if and only if the projection of $v$ on
$(c)tfspan(y_1,..., y_n)$ is zero. 
\end{dfn}

The concept of orthogonality is extended via the notion of Wiener separation, defined below. 

\begin{dfn}[Wiener separation \cite{materassi2019signal}]
Let $(H(z), e)$ be a LDIM with output processes 
$(y_{1},.,y_{n})$. 
We say that the process $y_{j}$ is Wiener separated from $y_{i}$ given a set of processes $S \subseteq y$ if the non-causal (causal) Wiener filter estimating $y_{j}$ from $y_{i} \cup S$ has a zero entry associated with $y_{i}$, 
i.e. $W_{y_{j}[y_{i}]|y_{i} \cup S} = 0$.
\end{dfn}

Depending on whether we use the non-causal or causal Wiener filter, we denote Wiener separation as 
$wsep(y_{j}, S, y_{i})$ and $cwsep(y_{j}, S, y_{i})$, respectively. 
Similarly, if Wiener separation does not hold, we use the notation
$\neg wsep(y_{j}, S, y_{i})$ and $\neg cwsep(y_{j}, S, y_{i})$.
We note that the special case when two processes $y_{i}$ and $y_{j}$ can be Wiener-separated with the empty set is equivalent to saying that $y_{i}$ and $y_{j}$ are orthogonal.

\section{Ill-posedness of the network reconstruction problem from observed data}\label{sec: counterexample false negatives}

In this section we show that the causal graph reconstruction problem is not well-posed, namely it does not admit, in general, a unique solution.
We provide a counterexample showing that without any additional assumptions on the network of the system, no method can guarantee an exact reconstruction of the skeleton of the network from observational data. 
One mechanism that we could use to show that the causal graph reconstruction problem is ill-posed, is by creating path cancellations that lead to the inference of a false negative edge.

\begin{figure}[h!]
	\centering
	\begin{tabular}{cc}
	\includegraphics[width=0.3\columnwidth]{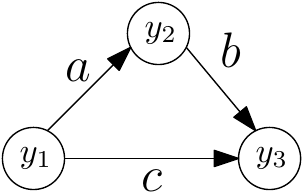} 
	&
	\includegraphics[width=0.3\columnwidth]{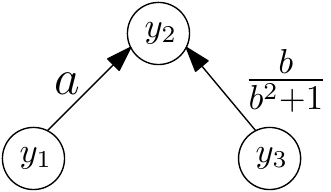}	
	\\ 
	(a) & (b) \\ 
	
	\end{tabular}\caption{(a) The causal graph of $\mathcal{G}_{1}$ (b) The causal graph of $\mathcal{G}_{2}$. We cannot distinguish between these two causal graphs given the dynamics of  $\mathcal{G}_{1}$ and  $\mathcal{G}_{2}$. \label{fig:triangle}}
\end{figure}

Consider a LDIM $\mathcal{G}_{1} = (H_{1}(z), E_{1})$ with transfer function 
\[H_{1}(z) = 
\begin{bmatrix}
0 & 0 & 0\\
a & 0 & 0 \\
c & b & 0
\end{bmatrix}  \,,
\quad \Phi_{E_{1}E_{1}}=\begin{bmatrix}
1 & 0 & 0\\
0 & 1 & 0 \\
0 & 0 & 1
\end{bmatrix}
\]
where $a,b,c \in \mathbb{R}$, $a,b,c \neq 0$ and output processes $Y_{1}$. The causal graph of this LDIM is shown in Figure~\ref{fig:triangle}(a).

Alternatively, consider the LDIM $\mathcal{G}_{2} = (H_{2}(z), E_{2})$:
\[H_{2}(z) = 
\begin{bmatrix}
0 & 0 & 0\\
a & 0 & \frac{b}{b^2+1} \\
0 & 0 & 0 
\end{bmatrix}; \quad \Phi_{E_{2}E_{2}} = 
 \begin{bmatrix}
1 & 0 & 0\\
0 & \frac{1}{b^2+1} & 0 \\
0 & 0 & b^2+1 
\end{bmatrix} \]
and output processes $Y_{2}$.
The causal graph of this LDIM is shown in Figure~\ref{fig:triangle}(b). 
Note for $c=-a\cdot b$ the PSDs of $\mathcal{G}_{1}$ and $\mathcal{G}_{2}$ are equal: 
 \[\Phi_{Y_{1}Y_{1}} = \Phi_{Y_{2}Y_{2}} =  
 \begin{bmatrix}
1 & a & 0\\
a & a^2+1 & b \\
0 & b & b^2+1 
\end{bmatrix} \,.\]

Notice that both LDIMs have no algebraic loops.
However, as the PSDs of both systems are the same, no method can distinguish these two systems from observational data only, thus no method can guarantee an exact reconstruction from observational data for networks with general dynamics and topology.  

The reason why these two systems cannot be distinguished is because of path cancellations occurring in $\mathcal{G}_{1}$. Namely, the influence of $y_{1}$ to $y_{3}$ through the direct edge $y_{1}\to y_{3}$ cancels the influence of $y_{1}$ to $y_{3}$ through the path $y_{1} \to y_{2} \to y_{3}$. Thus, in $\Phi_{Y_{1}Y_{1}}$ we get zeros in the entries $\{1,3\}$ and $\{3,1\}$. This enables us to construct another system, with a sparser causal graph, that has the same PSD as $\mathcal{G}_{1}$.

The ill-posedness of the problem leads to the question of how often can we have the same PSD for two systems with different causal graphs.
In order to answer that question, in the next section, we define the notion of Granger-faithful recursive LDIMs. 
We then provide a measure-theoretic analysis of Granger-unfaithful LDIMs. 

\section{Granger-faithfulness}\label{sec: GEMD and Granger-faithfulness}

In the previous section we showed that there are instances of recursive LDIMs for which an exact reconstruction is not possible.
Nonetheless, for recursive LDIMs, the Mixed-Delay (MD) method in \cite{dimovska2020control} can guarantee the absence of false positive edges in the reconstructed topology.
In this section we present a more-efficient variation of the MD algorithm in \cite{dimovska2020control} as it performs fewer tests for determining the existence of an edge in the topology. 
We call this variation Granger-Embedding Mixed Delay (GEMD) and we show that it has the same guarantees as the MD algorithm. 
Then, motivated by the steps of the GEMD algorithm, we formalize the notion of Granger-faithfulness and we analyze it in the Lebesgue measure sense. 

\subsection{Granger-Embedding Mixed Delay (GEMD) Algorithm}

The algorithm described in \cite{dimovska2020control} tries to Wiener separate two processes of a LDIM by exhaustively searching for a Wiener separating set among the following sets:
\begin{itemize}
    \item The set of all observed processes $y=\{y_{1},...,y_{n}\}$ union
    \item The set of all delayed processes denoted by $\frac{1}{z}y$
\end{itemize}
However, we show that the search space in the MD algorithm can be restricted to the processes in $y$ only, while preserving the same consistency guarantees.  delayed
To this end, we first state the Granger-Embedding variation of the MD algorithm that always considers all the processes in $\frac{1}{z}y$ (thus embedding Granger's method), and then we show that its consistency properties match the consistency properties of the MD method.

\subsection*{Granger-Embedding Mixed Delay (GEMD) Algorithm}
\begin{itemize}
 \item[{\tt}] Start with an empty graph on the vertices $V=\{y_{1}, ..., y_{n}\}$. For every unordered pair of nodes $\{ y_i, y_j \}$ perform the following three tests:
 \item[{\tt} 1. ] (Step 1: Feedthrough Wiener Separation) Search for a set $S_{ij}^{c} \subseteq y$, such that
 $W_{y_{j}[y_{i}]|\{ y_{i} \cup S_{ij}^{c} \cup \frac{1}{z}y \}}$ 
 is strictly causal.
 \item[{\tt}] \quad \quad  If the output of Step 1 is negative (no such set $S_{ij}^{c}$ is found), draw an undirected edge between $y_{i}$ and $y_{j}$.
 \item[{\tt} 2a. ] (Step 2a: Wiener separation of $y_{j}$ from $\frac{1}{z}y_{i}$) If a separating set was found in Step~1, continue to search for a set $S_{ji}^{d} \subseteq y$, such that 
 $W_{y_{j}[\frac{1}{z}y_{i}]|\{S_{ji}^{d}\cup \frac{1}{z}y \}} = 0$. 
\item[{\tt} 2b. ] (Step 2b: Wiener separation of $y_{i}$ from $\frac{1}{z}y_{j}$) If a separating set was found in Step~1, continue to search for a set $S_{ij}^{d} \subseteq y$, such that $W_{y_{i}[\frac{1}{z}y_{j}]|\{S_{ij}^{d}\cup \frac{1}{z}y \}} = 0$.
\item[{\tt}]    \quad \quad If a separating set was not found in either Step~2a or Step~2b, draw a double-headed link from $y_{i}$ to $y_{j}$ (Step~2a) or $y_{j}$ to $y_{i}$ (Step~2b).\\
\end{itemize}

Just as the output of the MD algorithm, the output of GEMD algorithm is a partially oriented multi-arrowed graph: it contains directed double-headed links which indicate a strictly causal influence only, and undirected edges which indicate the existence of a direct feedthrough between two variables. 

\begin{prop}\label{prop:GEMD consistency}
    Let $(H(z),e)$ be a recursive LDIM and let $G$ be its perfect graphical representation. 
    The output of GEMD does not infer any false positive edges, i.e. edges that are not present in $G$, and it matches the output of the MD algorithm. 
\end{prop}

\begin{IEEEproof}
 
    We inspect the proofs of Theorem~3.3 and Theorem~3.6 in \cite{dimovska2020control}. 
    Specifically, we notice that in the proofs of both of the theorems, we assume that we are given a recursive graphical representation of the input LDIM. 
    Without loss of generality, such a recursive graphical representation can have all the possible strictly causal influences. 
    Thus, the sets $S_{s}$ and $S^{-}$ (in Theorem~3.3 and Theorem~3.6 respectively), will be equal to $\frac{1}{z}y$, leading to the consistency of the GEMD steps. 

\end{IEEEproof}

\subsection{Granger-unfaithful LDIMs and their Lebesgue measure}

In the above algorithm, there are two steps, namely the feedthrough Wiener separation step, as well as the two steps that check for delayed dependencies between two nodes, that determine the existence of an edge from data. 
If one of those two steps finds a Wiener-separating set, while the corresponding edge exists in the graphical representation, we infer a false negative. 
Therefore, we could define a faithfulness condition that avoids such scenarios.
To that end, we first recall the standard notion of $d$-connection, which is one of the fundamental concepts in the theory of graphical models \cite{Pea88}. 
\begin{dfn}[D-connection]
Let $G=(V,E)$ be a directed graph, let $S \subseteq V$ and let $y_{i}, y_{j} \notin S$. We say that $y_{i}$ and $y_{j}$ are $d$-connected given $S$ if and only if: 
\begin{itemize}
      \item There exists an undirected path $U$ between $y_{i}$ and $y_{j}$ such that for every collider $C$ on $U$, either $C$ or a descendent of $C$ is in $S$ and
      \item No non-collider on $U$ is in $S$. 
\end{itemize}
\end{dfn}

In our framework we specialize the notion of $d$-connection to graphical representations. 
Since in a graphical representation we have two types of edges, we define a feedthrough and delayed $d$-connection. 

\begin{dfn}[Feedthrough D-connection]
Let $G=(V,E_{1}, E_{2})$ be a graphical representation of a LDIM and let $G^{\text{\Lightning}}$ be the corresponding graph of instantaneous propagations. 
Let $S \subseteq V$ and let $y_{i}, y_{j} \notin S$. We say that $y_{i}$ and $y_{j}$ are feedthrough $d$-connected in $G$ given $S \cup \frac{1}{z}y$ if and only if: 
\begin{itemize}
    \item  $y_{i}$ and $y_{j}$ are $d$-connected given $S$ in $G^{\text{\Lightning}}$. 
\end{itemize}
We denote the feedthrough $d$-connection of $y_{i}$ and $y_{j}$ given $S \cup \frac{1}{z}y$ as $d-conn(y_{j}, S \cup \frac{1}{z}y, y_{i})$. 
\end{dfn}

\begin{dfn}[Delayed D-connection]
Let $G=(V,E_{1}, E_{2})$ be a graphical representation of a LDIM and let $G^{\text{\Lightning}}$ be the corresponding graph of instantaneous propagations. 
Let $S \subseteq V$ and let $y_{i}, y_{j} \notin S$. We say that $y_{j}$ is delay $d$-connected with $y_{i}$ in $G$ given $S \cup \frac{1}{z}y$ if and only if: 
\begin{itemize}
     \item There is a double-headed edge from $y_{i}$ to some node, $y_{c}$ and 
    \item By substituting the outgoing double-headed edge from $y_{i}$ to $y_{c}$ with a single-headed one, $y_{i}$ and $y_{j}$ become feedthrough $d$-connected given $S \cup \frac{1}{z}y$. 

\end{itemize}
We denote the delay $d$-connection of $y_{j}$ with $y_{i}$ given $S \cup \frac{1}{z}y$ as $d-conn(y_{j}, S \cup \frac{1}{z}y, \frac{1}{z}y_{i})$.
\end{dfn}

\begin{figure}[h!]
	\centering
	\bgroup
	\begin{tabular}{c}
	\includegraphics[width=0.3\columnwidth]{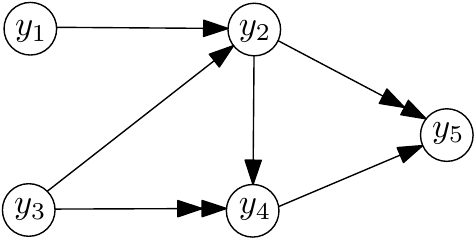}\\
	\includegraphics[width=0.8\columnwidth]{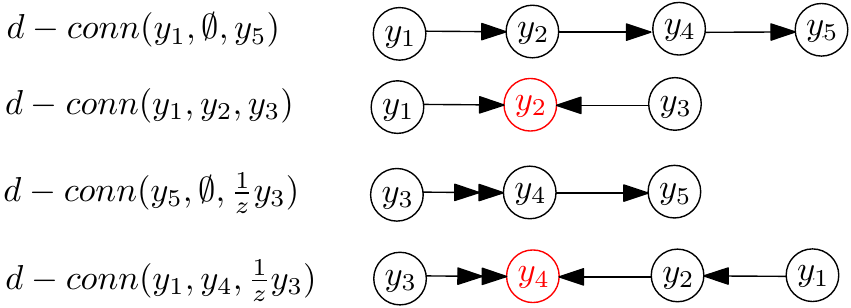}
	\end{tabular}
	\egroup
	\caption{ Examples of $d$-connection statements and the corresponding $d$-connecting paths in a graphical representation. Colliders that activate the $d$-connecting paths are shown in red. 
	\label{fig:dconnections} }
\end{figure}

We illustrate several $d$-connecting statements in a graphical representation in Figure~\ref{fig:dconnections}.

Next we formalize the notion of Granger-faithfulness which links the $d$-connection conditions to Wiener-separations.

\begin{dfn}[Granger-Faithfulness]
Let $\mathcal{G}=(H(z), e)$ be a recursive LDIM and let $G$ be a recursive graphical representation of the LDIM. 
We say the LDIM $\mathcal{G}$ is faithful to $G$ if for every ordered pair $(y_{i}, y_{j})$ the following two conditions are satisfied: 
\begin{itemize}
    \item $W_{y_{j}[y_{i}]|\{ S^{c} \cup \frac{1}{z}y \cup y_{i} \}}(\infty) \neq 0 \iff$  $y_{i}$ and $y_{j}$ are feedthrough $d$-connected given $S^{c} \cup \frac{1}{z}y$ in $G$ and 
    \item $W_{y_{j}[\frac{1}{z}y_{i}]|\{ S^{d} \cup \frac{1}{z}y \}}(z) \neq 0 \iff$ $y_{i}$ and $y_{j}$ are delay $d$-connected given $S^{d} \cup \frac{1}{z}y$ in $G$.
\end{itemize}
\end{dfn}

In order to show that Granger-unfaithfulness is only a pathological condition, we consider a class of recursive LDIMs $(H(z, \theta), e(\theta))$, parameterized by $\theta$. 
In particular, $\theta$ is the collection of coefficients of each of the real rational transfer function entries: 
\begin{equation}
    \begin{split}
    H_{ji}(z) &= \frac{b_{0}+b_{1}z^{-1}+...+b_{n_{1}}z^{-n_{1}}}{1+ a_{1}z^{-1}+...+a_{n_{2}}z^{-n_{2}}}\\
    F_{j}(z) &=  \frac{d_{0}+d_{1}z^{-1}+...+d_{n}z^{-n_{3}}}{1+ c_{1}z^{-1}+...+c_{n}z^{-n_{4}}}\,, \text{ where } e=F(z)u\,.
\end{split}
 \label{eqn:parameters of LDIM}
\end{equation}

\begin{thm}{(Zero-measure set)}\label{thm:zero-measure}
Let $\mathcal{G}$ be a class of LDIMs $(H(z, \theta), e(\theta))$ as in Equation~\ref{eqn:parameters of LDIM} with a recursive graphical representation $G$.
With respect to the Lebesgue measure over the space of parameters $\theta$, the set of LDIMs which are unfaithful to $G$ is a zero measure set. 
\end{thm}

\begin{IEEEproof}
 Here we just report a sketch of the proof. The proof with its full technical details is in Appendix~\ref{proof:zero-measure-details}.
 
 We consider all parametrized LDIMs for which $G$ is a graphical representation.
 Let 
 \begin{itemize}
    \item $d-conn(y_{j}, S^{d}\cup \frac{1}{z}y, \frac{1}{z}y_{i})$, or
    \item $d-conn(y_{j}, S^{c}\cup \frac{1}{z}y, y_{i})$ be true in $G$.
\end{itemize} 
The goal is to show that the set of parameters for which
\begin{equation}
    \begin{split}
    & W_{y_{j}[y_{i}]|S^{d}\cup \frac{1}{z}y}(z)\neq 0 \text{, if the delay $d$-conn. is true, or }\\
    & W_{y_{j}[y_{i}]|S^{c}\cup \frac{1}{z}y \cup y_{i}}(\infty)\neq 0 \text{, if the feedthrough $d$-conn. is true,}
\end{split}
 \label{eqn:wconn statements}
\end{equation}
has zero Lebesgue measure. 
As a first step we make use of a result in \cite{meek2013strong}, stating that there is a set of values 
of the LDIM parameters,
such that for a fixed $z=z_{\omega}=e^{i\omega}$ the statements in Relation (\ref{eqn:wconn statements}) are true. 

Since for a fixed $z$ the entries of the (causal) Wiener filter are polynomials (in this case in $\theta$),
the above results show that the $y_{i}$ entry in each of the statements in Relation~(\ref{eqn:wconn statements}) is a non-trivial polynomial. Indeed, if that entry were a trivial polynomial, then for all realizations of $\theta$, the $y_{i}$ component in the statements in Relation~(\ref{eqn:wconn statements}) would be zero. 
Therefore, the $y_{i}$ components in the expressions in Relation~(\ref{eqn:wconn statements}) are zero only for the values $\overline{\theta}$ that are in the solution set of the non-trivial polynomials in $\theta$. 
Then, we use a result from multivariate algebra that states that the set of solutions of a non-trivial polynomial is a zero measure set over the space of its parameters \cite{okamoto1973distinctness}. 
Using the completeness property of the Lebesgue measure, it follows that the set of values of $\theta$ for which the LDIMs with graphical representation $G$ are Granger-unfaithful to $G$ constitute a Lebesgue zero-measure set. 
\end{IEEEproof}
Therefore, assuming that a LDIM is Granger-faithful boils down to having a technical assumption which can be considered satisfied in most practical scenarios.

\section{Orientation inference in GEMD}\label{sec: orientations}

Under the condition of Granger-faithfulness, GEMD algorithm consistently recovers all double-headed links from the perfect graphical representation and produces an unoriented edge between $y_{i}$ and $y_{j}$ whenever there is a single-headed link between them in the perfect graphical representation of the LDIM.  
In what follows, we supplement the GEMD algorithm with additional orientation steps for the undirected edges, but first we give a brief overview of a graphical model algorithm that serves as the basis of the rules that we provide. 

\subsection{Orientation Inference in PC algorithm}

In order to orient the feedthroughs, we look at atemporal, graphical model techniques for inferring causal direction. 
In the area of graphical models, which in the linear case can be seen as LDIMs with coupling relations described only by direct feedthroughs, 
one of the most versatile structure identification tools is the PC algorithm \cite{spirtes1991algorithm}. 
Under the assumption of faithfulness, PC algorithm is able to recover the exact topology of a graphical model from observed data, by exhaustively testing for conditional independence of every pair of nodes $y_{i}$ and $y_{j}$ given a set $S \subseteq y$. 
In our LDIM framework, that is analogous to testing for the feedthrough Wiener-separation of $y_{i}$ and $y_{j}$ given $S \subseteq y$.
In addition, PC algorithm might also be capable of inferring some edge orientations as well.
We motivate how PC can orient edges by making use of an example where all the transfer functions are scalar gains.
\begin{example}\label{ldim:diamond orientations}
    Consider the LDIM with dynamics
    \begin{align*}
        &y_{i}=e_{i};            &y_{\ell}=y_{i}+e_{\ell}; &\qquad y_{j}=y_{\ell}+e_{j};\\
        &y_{k}=y_{i}+y_{j}+e_{k}; &y_{m}=y_{k}+e_{m} &`
    \end{align*}
    where the processes $e_{i}$, $e_{j}$, $e_{k}$, $e_{\ell}$, and $e_{m}$ are mutually independent and white
    and it is known that the transfer functions are scalar gains. The perfect graphical representation of this LDIM is shown in Figure~\ref{fig:pc_orientations}(a). 
    
    \begin{figure}[hb!]
	\centering
	\bgroup
	\def\arraystretch{0.6}
	\begin{tabular}{cc}
	\includegraphics[width=0.25\columnwidth]{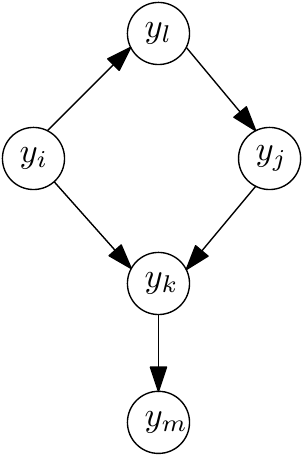} &
	\includegraphics[width=0.25\columnwidth]{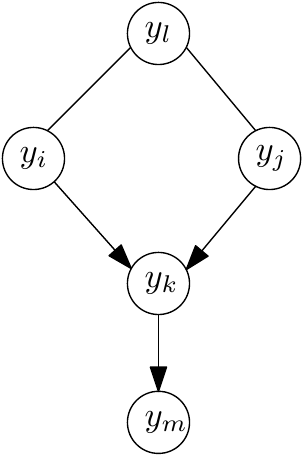}\\ 
    (a) & (b) \\
	\end{tabular}
	\egroup
	\caption{ (a) An LDIM with only direct feedthroughs as a motivating example demonstrating how PC algorithm discovers orientation of some edges of the inferred topology. (b) The reconstructed graphical representation of the LDIM in (a) by applying PC algorithm with orientation steps.
	\label{fig:pc_orientations} }
\end{figure}
        
    The LDIM is faithful, since every feedthrough $d$-connecting statement is observed via feedthrough Wiener separation from the data.
    Thus, using PC algorithm we can correctly infer the topology of the network.
    In particular, PC algorithm finds that $y_{i}$ and $y_{j}$ are not connected because, by choosing $S=\{y_{\ell}\}$ we obtain that $y_{i}$ and $y_{j}$ are conditionally independent given $S$.
    Now, since there is a path of length $2$ from $y_{i}$ and $y_{j}$ through the node $y_{k}$ and $y_{k}\notin S$, we can deduce the orientation of the links
    $y_{i}\to y_{k}$ and $y_{j}\to y_{k}$.
    In fact, we can rule out all other cases as follows:
    \begin{itemize}
        \item 
        If the orientations were
        $y_{j}\to y_{k} \to y_{i}$
        (a ``chain link'' from $y_{j}$ to $y_{i}$), then $y_{k}$ would be required to be in $S$ to ``block'' the information flow from $y_{j}$ to $y_{i}$.
        \item 
        If the orientations were
        $y_{i}\to y_{k} \to y_{j}$
        (a ``chain link'' from $y_{i}$ to $y_{j}$), then $y_{k}$ would be required to be in $S$ to ``block'' the information flow from $y_{i}$ to $y_{j}$.
        \item 
        If the orientations were
        $y_{i}\leftarrow y_{k}\to y_{j}$
        (a ``fork'' between $y_{i}$ and $y_{j}$), then $y_{k}$ would be required to be in $S$ to ``block'' the information  of $y_{k}$ shared in both $y_{i}$ and $y_{j}$. 
         
    \end{itemize}
    In the network under consideration, the only collider is $y_{i}\to y_{k}\leftarrow y_{j}$.
    After finding all the colliders in the graph, we can infer the orientation of additional edges.
    For example, it possible to deduce the orientation of the edge $y_{k}\to y_{m}$.
    Indeed, if the orientation was $y_{m}\to y_{k}$, we would have the collider configuration $y_{m}\to y_{k} \leftarrow y_{i}$ which was not found at the previous step.
\end{example}
 We note that PC in general cannot orient all edges. Indeed, in the LDIM in Example~\ref{ldim:diamond orientations}, PC is not able to infer orientations for the edges $y_{i} -y_{\ell}$ and $y_{j} - y_{\ell}$. Leaving those edges undirected indicates that they could be oriented in one of the three possible patterns: $y_{i} \rightarrow y_{\ell} \rightarrow y_{j}$; $y_{j} \rightarrow y_{\ell} \rightarrow y_{i}$; and $y_{i} \leftarrow y_{\ell} \leftarrow y_{j}$. A collider $y_{\ell}$ is ruled out. 
 The output of PC with the orientation steps is shown in Figure~\ref{fig:pc_orientations}(b).
 

\subsection{Orientation rules in the GEMD algorithm}\label{sec:orientation in MD}

Similarly to the orientation rules in the PC algorithm, we first attempt to detect the collider nodes in the output graph $G$ of the GEMD algorithm. 
As $G$ is a partially oriented graph, a collider node can be present between a double-headed and an undirected edge, and between two undirected edges.  
Next, we present two results that detect colliders in both of those cases.

\begin{prop}[Past to Present Collider Detection]\label{prop: orientation past to present}
    Let $(H(z, \theta),e(\theta))$ be a recursive, Granger-faithful LDIM and let $G$ be the output of the GEMD algorithm, when applied to $(H(z, \theta),e(\theta))$. Let the configuration
    $y_{i}$ \begin{tikzpicture}[font=\sffamily\tiny]
                \tikz \draw[>=triangle 60, ->>](0,-0.09) -- (0.7,-0.09);
            \end{tikzpicture} $y_{k}$
             \begin{tikzpicture}[font=\sffamily\tiny]
                \tikz \draw[-](0,-0.09) -- (0.7,-0.09);
            \end{tikzpicture} $y_{j}$
            be in $G$ and let a set $S_{ij}^{d}$ be found in Step~3.2a. of the GEMD algorithm, i.e. we can Wiener separate $y_{j}$ from $\frac{1}{z}y_{i}$. We have that the node $y_{k}$ is a collider between $y_{i}$ and $y_{j}$ if and only if $y_{k}\notin S_{ij}^{d}$. 
\end{prop}

\begin{IEEEproof}
 See Appendix~\ref{proof: orientation past to present}.  
\end{IEEEproof}

Next we present a similar result that detects collider nodes between two undirected edges.  

\begin{prop}[Feedthrough Collider Detection]\label{prop: orientation present to present}
 Let $(H(z, \theta),e(\theta))$ be a recursive, Granger-faithful LDIM and let $G$ be the output of the GEMD algorithm applied to the LDIM. Let the configuration
    $y_{i}$ \begin{tikzpicture}[font=\sffamily\tiny]
                \tikz \draw[-](0,-0.09) -- (0.7,-0.09);
            \end{tikzpicture} $y_{k}$
             \begin{tikzpicture}[font=\sffamily\tiny]
                \tikz \draw[-](0,-0.09) -- (0.7,-0.09);
            \end{tikzpicture} $y_{j}$
            be in $G$ and let $S_{ij}^{c}$, a Wiener separating set for $y_{i}$ and $y_{j}$ be found in step $3.1$  of the GEMD algorithm. The node $y_{k}$ is a collider between $y_{i}$ and $y_{j}$ if and only if $y_{k}\notin S_{ij}^{c}$. 
\end{prop}

\begin{IEEEproof}
 See Appendix~\ref{proof: orientation present to present}. 
\end{IEEEproof}

The results of Proposition~\ref{prop: orientation past to present} and Proposition~\ref{prop: orientation present to present} help us detect two types of colliders. After finding all the colliders, we can then further infer some link orientations related to nodes that have not been found to be colliders, but have to be chain links. 
The orientation deducing steps that we apply to the output of the GEMD algorithm are summarized below. 

\begin{itemize}
    \item   
    {\bf 1. Detection of all colliders}. \\
    For every undirected edge between $y_i$ and $y_k$ in the output graph by GEMD, apply the following steps
    \begin{itemize}
        \item[\tiny$\bullet$]
        {\bf Type~A (past-present-present) collider, namely}\\
            {\bf  ``$y_j$ 
            \begin{tikzpicture}[font=\sffamily\tiny]
                \tikz \draw[>=triangle 60, ->>](0,-0.09) -- (0.7,-0.09);
            \end{tikzpicture}
            $y_k$
            \begin{tikzpicture}[font=\sffamily\tiny]
                \tikz \draw[>=triangle 60, <-](0,-0.09) -- (0.7,-0.09);
            \end{tikzpicture}
            $y_{i}$''}\\
        If the edge ``$y_j$ 
		\begin{tikzpicture}[font=\sffamily\tiny]
			\tikz \draw[>=triangle 60, ->>](0,-0.09) -- (0.7,-0.09);
		\end{tikzpicture} $y_k$'' 
		is in the graph, a set $S^{d}_{ij}$ was found by GEMD, 
        and $y_k$ is not an element of the set $S^{d}_{ij}$, then orient the edge as 
        ``$y_i$ 
        \begin{tikzpicture}[font=\sffamily\tiny]
            \tikz \draw[>=triangle 60, ->](0,-0.09) -- (0.5,-0.09);
        \end{tikzpicture} $y_k$''
        
        \item[\tiny$\bullet$]
        {\bf Type~B (present-present-present) collider, namely}\\
            {\bf  ``$y_j$            
            \begin{tikzpicture}[font=\sffamily\tiny]
                \tikz \draw[-](0,-0.09) -- (0.7,-0.09);
            \end{tikzpicture}
            $y_k$
            \begin{tikzpicture}[font=\sffamily\tiny]
                \tikz \draw[-](0,-0.09) -- (0.7,-0.09);
            \end{tikzpicture}
            $y_{i}$''}\\
        If the edge ``$y_j$ 
		\begin{tikzpicture}[font=\sffamily\tiny]
			\tikz \draw[-](0,-0.09) -- (0.7,-0.09);
		\end{tikzpicture} $y_k$'' 
		is in the graph, a set $S^{c}_{ij}$ was found by GEMD, 
        and $y_k$ is not an element of the set $S^{c}_{ij}$, then orient
        ``$y_j$ 
        \begin{tikzpicture}[font=\sffamily\tiny]
            \tikz \draw[>=triangle 60, ->](0,-0.09) -- (0.5,-0.09);
        \end{tikzpicture} $y_k$
        \begin{tikzpicture}[font=\sffamily\tiny]
            \tikz \draw[>=triangle 60, <-](0,-0.09) -- (0.5,-0.09);
        \end{tikzpicture} $y_{i}$''. 
        
    \end{itemize}
    \item 
    {\bf 2. Propagation of orientations}. \\
    After the ``detection of all colliders'' step, for every edge left undirected $(y_{i}, y_{j})$ do the following: 
    \begin{itemize}
         \item[\tiny$\bullet$]
        {\bf Type~A propagation} \\
        If $y_{i}$ 
        \begin{tikzpicture}[font=\sffamily\tiny]
            \tikz \draw[-](0,-0.09) -- (0.7,-0.09);
        \end{tikzpicture}
            $y_{j}$ is adjacent to 
        ``$y_k$ 
        \begin{tikzpicture}[font=\sffamily\tiny]
			\tikz \draw[>=triangle 60, ->>](0,-0.09) -- (0.7,-0.09);
		\end{tikzpicture}
        $y_i$''
        and a set $S^{d}_{jk}$ was found by GEMD, orient
        $y_i$
         \begin{tikzpicture}[font=\sffamily\tiny]
            \tikz \draw[>=triangle 60, ->](0,-0.09) -- (0.5,-0.09);
        \end{tikzpicture} $y_j$''.
         \item[\tiny$\bullet$]
          {\bf Type~B propagation} \\
        If $y_{i}$         
        \begin{tikzpicture}[font=\sffamily\tiny]
            \tikz \draw[-](0,-0.09) -- (0.7,-0.09);
        \end{tikzpicture}
        $y_{j}$ is adjacent to 
        ``$y_k$ 
        \begin{tikzpicture}[font=\sffamily\tiny]
			\tikz \draw[>=triangle 60, ->](0,-0.09) -- (0.7,-0.09);
		\end{tikzpicture}
        $y_i$''
        and a set $S^{c}_{jk}$ was found by GEMD, orient
        $y_i$
         \begin{tikzpicture}[font=\sffamily\tiny]
            \tikz \draw[>=triangle 60, ->](0,-0.09) -- (0.5,-0.09);
        \end{tikzpicture} $y_j$''.
    \end{itemize}
\end{itemize}

\begin{figure}[hb!]
	\centering
	\begin{tabular}{cc}
	\includegraphics[width=0.3\columnwidth]{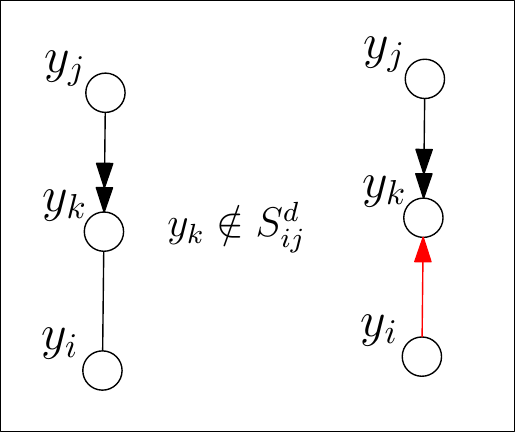} &
	\includegraphics[width=0.3\columnwidth]{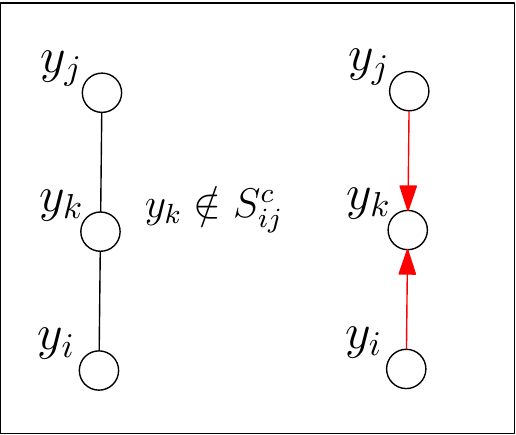} \\
	 (Type~A collider) & (Type~B collider) \\
        ~&~\\
	\includegraphics[width=0.3\columnwidth]{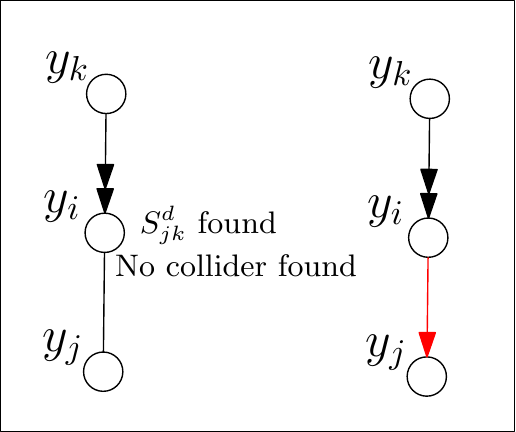} &
	\includegraphics[width=0.3\columnwidth]{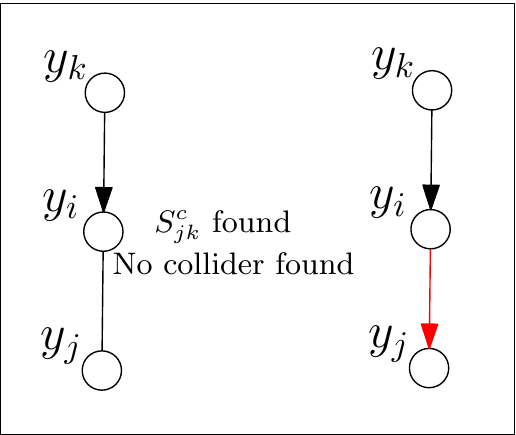} \\
	 (Type~A propagation) & (Type~B propagation)
	\end{tabular}
	\caption{For each subfigure, left: the partially oriented graph $G$, output of GEMD; and right: the inferred direction in $G$ after applying the orientation steps. \label{fig:orientation rules}}
\end{figure}

~\\
Thus, the orientation rules for GEMD algorithm are essentially a search for all colliders, followed by the propagation of additional orientations. 
Figure~\ref{fig:orientation rules} explains the process. 

We illustrate these orientation rules through an example. 

\begin{example}\label{ldim:md}
Consider a LDIM with perfect graphical representation shown in Figure~\ref{fig:md example ldim}a. 

\begin{figure}[hb!]
	\centering
	\bgroup
	\def\arraystretch{0.6}
	\begin{tabular}{cc}
  
	\includegraphics[width=0.45\columnwidth]{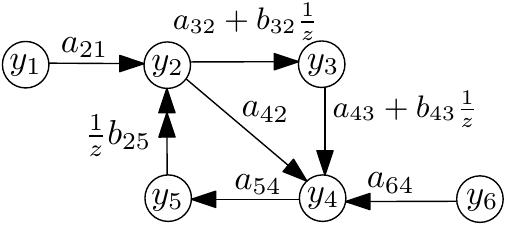} &
	 \includegraphics[width=0.45\columnwidth]{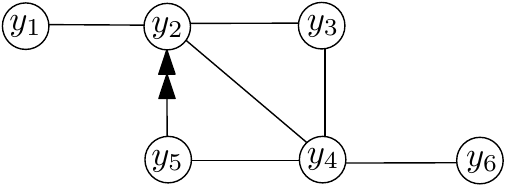} \\ 
    
	& \\
    (a) & (b)
	\end{tabular}
	\egroup
	\caption{ (a) Perfect graphical representation of a recursive LDIM  (faithfulness assumed). The transfer functions are given on the links. (b) The reconstructed partially oriented network after applying the GEMD algorithm (before applying orientation rules). \label{fig:md example ldim} }
\end{figure}

Note that this LDIM features a feedback loop, colliders, feedthrough links and a strictly causal link. 
Despite the complex elements in this LDIM, using the GEMD algorithm we can recover the exact topology of the network, together with the double-headed edge from $y_{5}$ to $y_{2}$, as long as the LDIM is Granger-faithful to its perfect graphical representation. The theoretical output of GEMD for this LDIM is shown in Figure~\ref{fig:md example ldim}b.
We demonstrate this empirically as well, by simulating the LDIM for $1000$ times. 
In each simulation we choose the values for the parameters $a_{21}, a_{32}, a_{42}, a_{43}, a_{54}, a_{64}, b_{32}, b_{25}$ uniformly at random from the interval $(0.3, 0.6)$. 
In each simulation, we run the GEMD algorithm and for each $(j, i)$ pair  we obtain an $f$-score for the existence of a direct link between $(y_{j}, y_{i})$ and $(y_{j}, \frac{1}{z}y_{i})$. 
Intuitively, the $f$-score tells us what is the maximum reduction in error variance that we can achieve when predicting $y_{j}$ by considering $y_{i}$ (or $\frac{1}{z}y_{i}$) as a parent of $y_{j}$, given a set $S \cup \frac{1}{z}y$.  
If this reduction is greater than a threshold $\theta \in [0,1]$, then $y_{i}$ (or $\frac{1}{z}y_{i}$) would be inferred as parent of $y_{j}$. 
More details regarding the calculation of the $f$-score can be found in \cite{dimovska2020control}, where a similar simulations-based approach was taken with the objective of reconstructing only the topology of a recursive LDIM. 
We record the $f$-scores from each such pair in each simulation and we then use them to obtain an ROC curve. 
Since in this case we want to compare the obtained $f$-scores against the partially directed graph shown in Figure~\ref{fig:md example ldim}b, the true scores are given by concatenating together the adjacency matrix of the graph of instantaneous propagations, and the adjacency matrix obtained by the strictly causal connections. 
The ROC curves obtained by those simulations, 
given different data horizons,
are shown in Figure~\ref{fig:GEMD ROC}. As we can see, the empirical output of GEMD closely matches the theoretical one, even for small data sizes. 

\begin{figure}[hb!]
	\centering
	\bgroup
	\def\arraystretch{0.8}
	\begin{tabular}{c}
  
	\includegraphics[scale=0.4]{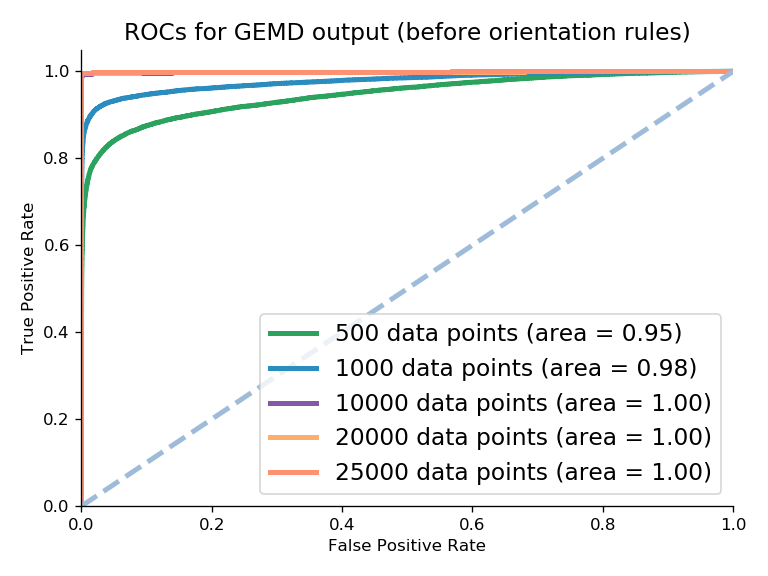} 
	\end{tabular}
	\egroup
	\caption{ ROCs from simulations evaluating the output of GEMD on the input LDIM from Example~\ref{ldim:md}, before applying orientation rules.
	\label{fig:GEMD ROC} }
\end{figure}

Next, we demonstrate the orientation steps of the GEMD method. In this example, it happens that we can recover all the orientations, however, we note that that is not always the case. 
The detailed steps for inferring the orientations are as follows: 
\begin{itemize}
  \item Orient $y_{1} - y_{2}$: (Type~A collider detection) The variable $y_{2} \notin S^{d}_{51}$. Thus, $y_{1}$
    \begin{tikzpicture}[font=\sffamily\tiny]
		\tikz \draw[>=triangle 60, ->](0,-0.09) -- (0.7,-0.09);
	\end{tikzpicture} $y_{2}$. 
	
    \item Orient $y_{3} - y_{4}$ and $y_{4}-y_{6}$: (Type~B collider) A set $S^{c}_{36}$ s.t. $y_{4} \notin S^{c}_{36}$is found. Thus, $y_{3}$ 
	\begin{tikzpicture}[font=\sffamily\tiny]
        \tikz \draw[>=triangle 60, ->](0,-0.09) -- (0.7,-0.09);
    \end{tikzpicture} $y_{4}$ and  $y_{6}$ 
	\begin{tikzpicture}[font=\sffamily\tiny]
        \tikz \draw[>=triangle 60, ->](0,-0.09) -- (0.7,-0.09);
    \end{tikzpicture} $y_{4}$. 	
    
\item Orient $y_{2}-y_{4}$: (Type~A collider) A set $S^{c}_{26}$ was found and $y_{4} \notin S^{c}_{26}$. Thus, $y_{2}$ 
	\begin{tikzpicture}[font=\sffamily\tiny]
        \tikz \draw[>=triangle 60, ->](0,-0.09) -- (0.7,-0.09);
    \end{tikzpicture} $y_{4}$. 
    
      \item Orient $y_{2} - y_{3}$:  (Type~A propagation) The variable $y_{2} \in S^{d}_{35}$. Thus, $y_{2}$ 
	\begin{tikzpicture}[font=\sffamily\tiny]
        \tikz \draw[>=triangle 60, ->](0,-0.09) -- (0.7,-0.09);
    \end{tikzpicture}	$y_{3}$. Note that Type~B propagation would also find the same orientation. 
          			
    \item Orient $y_{4} - y_{5}$: (Type~B propagation) The set $S^{c}_{35}$ has been found by GEMD and $y_{4}$ has not been detected as a collider between $y_{3}$ and $y_{5}$ in previous orientation steps. Thus, $y_4$ 
	\begin{tikzpicture}[font=\sffamily\tiny]
        \tikz \draw[>=triangle 60, ->](0,-0.09) -- (0.7,-0.09);
    \end{tikzpicture} $y_{5}$.
    
\end{itemize}

The above steps are confirmed empirically as well.
From the ROC curves shown in Figure~\ref{fig:GEMD ROC}, in each simulation horizon we obtain the threshold for which we get the highest number of true positive and the lowest number of false positive inferred edges. Based on that threshold, in each simulation we first reconstruct the skeleton and then we apply the orientation inference rules.  
Namely, in each simulation, for each pair, we also record the set that gave rise to the minimum $f$-score, which we regard as the Wiener-separating set. 
Then, for every triple, $(y_{i}, y_{k}, y_{j})$ we first apply the collider detection rules, orienting every Type~A and Type~B collider triple. 
Then, for each unoriented edge, we check if an orientation propagation rule can be applied. 
Once we orient all the edges, we compare the inferred orientation for each edge with its true orientation, obtaining the accuracy of the orientation rules across all the simulations. 
The accuracy across all simulations, for different data horizons, is given in Table~\ref{tab:orientations accuracy}. 
\begin{center}
    \begin{table}\caption{Accuracy of orientation rules
	\label{tab:orientations accuracy}}
    \centering
        \begin{tabular}{|l|l|l|l|l|l|}
        \hline
        Data points & 500  & 1000 & 10000 & 20000 & 25000\\ \hline
        Accuracy    & 0.81 & 0.86 & 0.96  & 0.98 & 0.99  \\ \hline
        \end{tabular}
    
    \end{table}
\end{center}
\end{example}

\section{Conclusion}\label{sec:conclusion}

This article elaborated on the problem of network reconstruction from non-invasively observed data in the presence of direct feedthroughs and feedback loops.
Many existing methods have various limitations: some of them infer both false positive and false negatives edges in the presence of feedthroughs and/or feedback loops, while few methods provide guarantees for no false positives only, even when direct feedthroughs are present. 
The case of false negatives is more delicate: as we show in this article, false negatives can arise in pathological situations, occurring under specific values of the system parameters.
In such cases, borrowing a terminology from the theory of graphical models, we say that the network is not ``faithful.''
In this article we prove that positing the faithfulness of a network is a mild assumption since it is equivalent to excluding pathological situations that happen on zero-measure sets. 
In particular, we define a variation of faithfulness for linear dynamic systems, called Granger-faithfulness 
and we show that the systems that are Granger-unfaithful have Lebesgue zero measure. 
Under the assumption of Granger-faithfulness, we provide an algorithm for network reconstruction that guarantees the inference of the true network topology, with no false positive nor false negative edges being inferred in the limit of infinite data.
The algorithm also correctly infers all the links with strictly casual influences and orients them. 
Lastly, assuming Granger-faithfulness, we provide a set of rules that can determine the orientation of some edges associated with direct feedthroughs as well. 


\section{Appendix}

\subsection{Proof of Theorem~\ref{thm:zero-measure}}

\begin{IEEEproof}\label{proof:zero-measure-details}
 
Since faithfulness of LDIMs consists of two conditions, we need to consider these two $d$-connecting statements:
\begin{itemize}
    \item Let the delay $d$-connecting statement, $d-conn(y_{j}, S^{d}\cup \frac{1}{z}y, \frac{1}{z}y_{i})$, be true in $G$, or
    \item Let the feedthrough $d$-connecting statement $d-conn(y_{j}, S^{c}\cup \frac{1}{z}y, y_{i})$ be true in $G$.
\end{itemize} 

In order to connect the Wiener separating statements that correspond to the above $d$-connecting statements, we begin by considering the parametric expression of both the non-causal and the causal Wiener filter for estimating $y_{j}$ from an observed process $y_{i}$ given a set $S \subseteq \{y \setminus \{y_{i}, y_{j}\}) \}$.

We first consider the non-causal Wiener filter. 
\begin{align*}
 &W^{nc}_{y_{j}[y_{i}]|S}(z, \theta) = \\
 &[\Phi_{ji}(z, \theta) \quad \Phi_{jS}(z, \theta) ]
                \begin{pmatrix} \Phi_{ii}(z, \theta) & \Phi_{iS}(z, \theta) \\
                              \Phi_{S i}(z, \theta) & \Phi_{SS}(z, \theta)\end{pmatrix}^{-1}
                                \begin{pmatrix} 1  \\
                              \vec{\textbf{0}} \end{pmatrix} \,.
\end{align*}

Regarding the causal Wiener-Filter expression, we need to consider the spectral factorization \cite{sayed2001survey}

\[
 \begin{pmatrix} \Phi_{ii}(z, \theta) & \Phi_{iS}(z, \theta) \\
                              \Phi_{S i}(z, \theta) & \Phi_{SS}(z, \theta)\end{pmatrix} = F(z, \theta)F(z, \theta)^{*}\,,
\]
where $F(z, \theta)$ is a lower-triangular matrix with real-rational entries $\frac{K(z, \theta)}{L(z, \theta)}$ with no zeros or poles on the unit circle. 

Using Wiener-Kinchin theorem we have that:

\begin{align*}
 &W^{c}_{y_{j}[y_{i}]|S}(z, \theta) = \\
 &\mathcal{C}\left\lbrace[\Phi_{ji}(z, \theta) \quad \Phi_{jS}(z, \theta) ]F^{-*}(z, \theta)\right\rbrace
 F^{-1}(z, \theta) \begin{pmatrix} 1  \\
                              \vec{\textbf{0}} \end{pmatrix} \,,
\end{align*}
where $\mathcal{C}$ denotes the truncation of each entry in the vector expression to a causal transfer function. 

Thus, in both the causal and non-causal Wiener filter expression (for any $S$), the $y_{i}$ component is a real rational function of $z$ with parametric expression
$\frac{P(z, \theta)}{Q(z, \theta)} = \frac{\sum_{k=0}^{m}{p_{k}(\theta)z^{-k}}}{\sum_{k=0}^{n}{q_{k}(\theta)z^{-k}}}$. 
Now, consider the expression $\frac{P(z, \theta)}{Q(z, \theta)}$ for a fixed $z=z_{\omega} = e^{i\omega}$. 
Then, the numerator $\mathcal{N_{\omega}}(\theta)$ of $\frac{P(z_{\omega}, \theta)}{Q(z_{\omega}, \theta)}$ is a multivariate polynomial in $\theta$ with complex coefficients. 
We will show that $\mathcal{N_{\omega}}(\theta)$ is a non-trivial polynomial in $\theta$, i.e. its coefficients are not identically zero.
If $\mathcal{N_{\omega}}(\theta)$ were identically zero, that means that in all possible parametrizations of $H$ (for which $G$ is a graphical representation),  $wsep(y_{j}, S, y_{i})$ holds. 
Therefore, in order to show that $\mathcal{N_{\omega}}(\theta)$  is non-trivial, we will construct realizations of $H(e^{i\omega}, \theta), e(\theta)$, for which $G$ is a graphical representation, and 
such that 
\begin{itemize}
    \item $\neg wsep(y_{j}, S\cup \frac{1}{z}y, \frac{1}{z}y_{i})$, if the delay $d$-connecting statement is true, or
    \item $W_{y_{j}[y_{i}]|S\cup \frac{1}{z}y \cup y_{i}}(\infty)\neq 0$, if the feedthrough $d$-connecting statement is true.
\end{itemize} 

We have the following cases: 

\begin{enumerate}
    \item If $y_{i}$
            \begin{tikzpicture}[font=\sffamily\tiny]
                \tikz \draw[>=triangle 60, ->](0,-0.09) -- (0.7,-0.09);
            \end{tikzpicture} $y_{j}$ or 
            $y_{i}$\begin{tikzpicture}[font=\sffamily\tiny]
                \tikz \draw[>=triangle 60, ->>](0,-0.09) -- (0.7,-0.09);
            \end{tikzpicture} $y_{j}$ 
             in $G$, then we construct $H$ such that $H_{kl}= 0$ for all $k,l\neq i,j$ and $H_{ji}=1$ (in the feedthrough $d$-connection case) or
             $H_{ji}=\frac{1}{z}$ (in the delay $d$-connection case). 
             We place the parameters in $F(e^{i\omega},\theta)$ such that $\Phi_{ee}$ is the identity matrix, i.e. $F_{j}(e^{i\omega})=1$, for all $j=1, ..., n$. 
             With this we reduce the perfect graphical representation of the LDIM to just the edge from $y_{i}$ to $y_{j}$. 
            In this case, the corresponding $\neg wsep$ statements hold trivially and $G$ remains a graphical representation of the LDIM. 
            
        \item If $y_{i}$
            \begin{tikzpicture}[font=\sffamily\tiny]
                \tikz \draw[>=triangle 60, ->](0,-0.09) -- (0.7,-0.09);
            \end{tikzpicture} $y_{j}$ 
            is not in $G$ but there is a feedthrough $d$-connecting path between $y_{i}$ and $y_{j}$, then we consider the $d$-connecting path properties. 
            Recall that by definition, this means that there is a $d$-connecting path  $p^{\text{\Lightning}}$ (given $S$), between $y_{i}$ and $y_{j}$ that lies in $G^{\text{\Lightning}}$.
            
    Now we use a result presented in \cite{geiger1993logical} that states that if two nodes are $d$-connected in a DAG, then there exists a minimal polytree subgraph of the DAG that contains the $d$-connecting path between $y_{i}$ and $y_{j}$. 
    Therefore, the nodes of $p^{\text{\Lightning}}$ lie on a polytree $G'$, since $G^{\text{\Lightning}}$ is a DAG.
    Similarly, in Lemma~11 in \cite{meek2013strong} it is shown that for two nodes that are $d$-connected given $S$ via a path that lies on a polytree, there is a positive Gaussian probability distribution such that  
    those two nodes are also correlated given $S$. 
    In other words, using the result in Lemma~11 from \cite{meek2013strong} we have that there are values of the parameters in $\theta$, namely a realization $\theta'$ such that $\Phi_{yy}$ is constant, the non-zero values in $\theta'$ are the ones associated with the links in the polytree $G'$ and: 
    
       \begin{align*}
        \hat y_{j} &=
        \operatorname*{arg\,min}_{q \in ctfspan(S \cup \frac{1}{z}y \cup y_{i})}{\left\| y_{j} - q \right\|^{2} } = \\
        & = \alpha_{i}y_{i} + \sum_{y_{l} \in S} G_{jl}y_{l} + F_{j}\frac{1}{z}e_{j} \,,
    \end{align*}

where $\alpha_{i} \neq 0$.
Note that, since $\Phi_{yy}$ is constant, $ctfpan(S \cup \frac{1}{z}y \cup y_{i}) = ctfspan(S \cup y_{i})$. 
Therefore, $W_{y_{j}[y_{i}]| S \cup \frac{1}{z}y \cup y_{i}}(\infty) \neq 0$, i.e. $\neg wsep(y_{j}, S\cup \frac{1}{z}y, y_{i})$ is true, while $G$ is still a graphical representation of the LDIM. 

\item The third case is when
          $y_{i}$
            \begin{tikzpicture}[font=\sffamily\tiny]
                \tikz \draw[>=triangle 60, ->>](0,-0.09) -- (0.7,-0.09);
            \end{tikzpicture} $y_{j}$ 
            is not in $G$ but
            there is a delayed $d$-connecting path from $y_{i}$ to $y_{j}$. 
            Recall that by definition, on the delayed $d$-connecting path from $y_{i}$ to $y_{j}$ there is a double-headed child of $y_{i}$ in $G$, namely $y_{c}$, such that if we replace 
            the double-headed edge from $y_{i}$ to $y_{c}$ with a single-headed one, we have that $y_{i}$ and $y_{j}$ become feedthrough $d$-connected given $S$. 
            Let us consider the delay $d$-connecting path between 
            $y_{i}$ and $y_{j}$ after replacing the double-headed edge from $y_{i}$ to $y_{c}$ with a single-headed edge. 
            Due to this edge transformation, let us change the notation of the node $y_{i}$ and instead let us denote it with $y_{i}'$.  
            We denote the (feedthrough) $d$-connecting path between $y_{i}'$ and $y_{j}$ by $p^{\text{\Lightning}}$. 
            As in the previous case, we know that  $p^{\text{\Lightning}}$ lies in $G^{\text{\Lightning}}$, thus $p^{\text{\Lightning}}$ lies on a minimal polytree $G'$, where $G'$ is a subgraph of $G$. 
             Further there is a positive Gaussian distribution that gives rise to a realization $\theta'$, such that 
    \begin{align*}
        \hat y_{j} &=
        \operatorname*{arg\,min}_{q \in ctfspan(S \cup \frac{1}{z}y \cup y_{i}')}{\left\| y_{j} - q \right\|^{2} } = \\
        & = \alpha_{i}'y_{i}' + \sum_{y_{l} \in S} G_{jl}y_{l} + F_{j}\frac{1}{z}e_{j} \,,
    \end{align*}
    with $\alpha_{i}'\neq 0$. 
    
 Note that, in $\theta'$, there is a non-zero coefficient $\alpha_{ci}'$ such that $y_{c} = \alpha_{ci}'y_{i}'$, since we know that $y_{c}$ is a child of $y_{i}$ and the probability distribution is positive.
 Now, we use the fact that in the LDIM with graphical representation $G'$, $y_{i} = zy_{i}'$, i.e. $y_{i}'=\frac{1}{z}y_{i}$. 
 Therefore, in $G'$ we can replace the  outgoing edge from $y_{i}$ to $y_{c}$ to a  double-headed edge, i.e.
    $y_{i}$ \begin{tikzpicture}[font=\sffamily\tiny]
                \tikz \draw[>=triangle 60, ->>](0,-0.09) -- (0.7,-0.09);
            \end{tikzpicture} $y_{c}$. 
Note that the realization of the parameters, namely $\theta'$, stays the same.
    
Therefore, there is a realization $\theta'$, with perfect graphical representation $G'$ and graphical representation $G$ such that 
    $W_{y_{j}[\frac{1}{z}y_{i}]| S \cup \frac{1}{z}y} \neq 0$, i.e. $\neg wsep(y_{j}, S\cup \frac{1}{z}y, \frac{1}{z}y_{i})$.
 
\end{enumerate}

Thus, $\mathcal{N_{\omega}}(\theta)$  is a non-trivial polynomial in $\theta$ with complex coefficients for both wsep statements. 
Therefore, in order for those wsep statements to be zero, we necessarily need a realization $\bar\theta$ s.t. $\bar \theta$ is in the solution set of the non-trivial polynomial $\mathcal{N_{\omega}}(\theta)$. 

Now we use the result that the set of solutions of a non-trivial polynomial is a zero measure set over the space of its parameters \cite{okamoto1973distinctness}.
Thus the set of values $\overline\theta$ for which $\mathcal{N_{\omega}}(\theta) = 0$,
denoted as $\mathcal{R}(z_{\omega})$,
constitutes a zero-measure set w.r.t. the Lebesgue measure over $\mathcal{C}^{nxn}$, i.e. $\lambda(\mathcal{R} (z_{0})) = 0$. 

Note that unfaithfulness means that
  $W^{c}_{y_{j}[y_{i}]|S}(z, \theta)$ and $W^{nc}_{y_{j}[y_{i}]|S}(z, \theta)$ are zero for all $z$ on the unit disk. 
  Thus, the LDIM is unfaithful for the set of values of the parameters $\theta$ that are in the set 
  $\mathcal{R}(z)= \bigcap\limits_{z_{i}} \mathcal{R}(z_{i}) $. By the completeness of the Lebesgue measure, $\lambda(\mathcal{R}(z))=0$.

\end{IEEEproof}

\subsection{Proof of Proposition ~\ref{prop: orientation past to present}}

\begin{IEEEproof}\label{proof: orientation past to present}
In the partially oriented graph $G$ (output of GEMD) we have  $y_{i}$ \begin{tikzpicture}[font=\sffamily\tiny]
                \tikz \draw[>=triangle 60, ->>](0,-0.09) -- (0.7,-0.09);
            \end{tikzpicture} $y_{k}$
             \begin{tikzpicture}[font=\sffamily\tiny]
                \tikz \draw[-](0,-0.09) -- (0.7,-0.09);
            \end{tikzpicture} $y_{j}$.
We also know that a set $S_{ji}^{d}$ has been found. 
Note that this means that first a set $S_{ji}^{c}$ was found, since the step that searches for a delay separating set is only executed after a feedthrough separating set has been found. 
Therefore, since the LDIM is Granger-faithful, we know that 
\begin{itemize}
    \item $y_{i}$ and $y_{j}$ are not feedthrough $d$-connected, and further
    \item $y_{j}$ is not delay $d$-connected with $y_{i}$ 
\end{itemize}
Since $y_{j}$ is not delay $d$-connected with $y_{i}$, this means that all the delay $d$-connecting paths from $y_{i}$ to $y_{j}$ are blocked. If $y_{k}$ was a collider on the path 
$y_{i}$ \begin{tikzpicture}[font=\sffamily\tiny]
                \tikz \draw[>=triangle 60, ->>](0,-0.09) -- (0.7,-0.09);
            \end{tikzpicture} $y_{k}$
             \begin{tikzpicture}[font=\sffamily\tiny]
                \tikz \draw[-](0,-0.09) -- (0.7,-0.09);
            \end{tikzpicture} $y_{j}$, then this path would
not be a delay $d$-connecting path given $S_{ji}^{d}$ if and only if $y_{k} \notin S_{ji}^{d}$. Otherwise, $y_{j}$ would be delay $d$-connected with $\frac{1}{z}y_{i}$ given $S_{ji}^{d}$, which is a contradiction since the LDIM is Granger-faithful. 
\end{IEEEproof}

\subsection{Proof of Proposition ~\ref{prop: orientation present to present}}

\begin{IEEEproof}\label{proof: orientation present to present}
 
The proof is analogous to the proof of Proposition~\ref{prop: orientation past to present} and basically follows from the definition of feedthrough $d$-connection and the properties of Granger-faithful LDIMs. 
\end{IEEEproof}

\bibliography{trees}

\vspace{-0.3in}
\begin{IEEEbiography}[{\includegraphics[width=1in,height=1.25in,keepaspectratio]{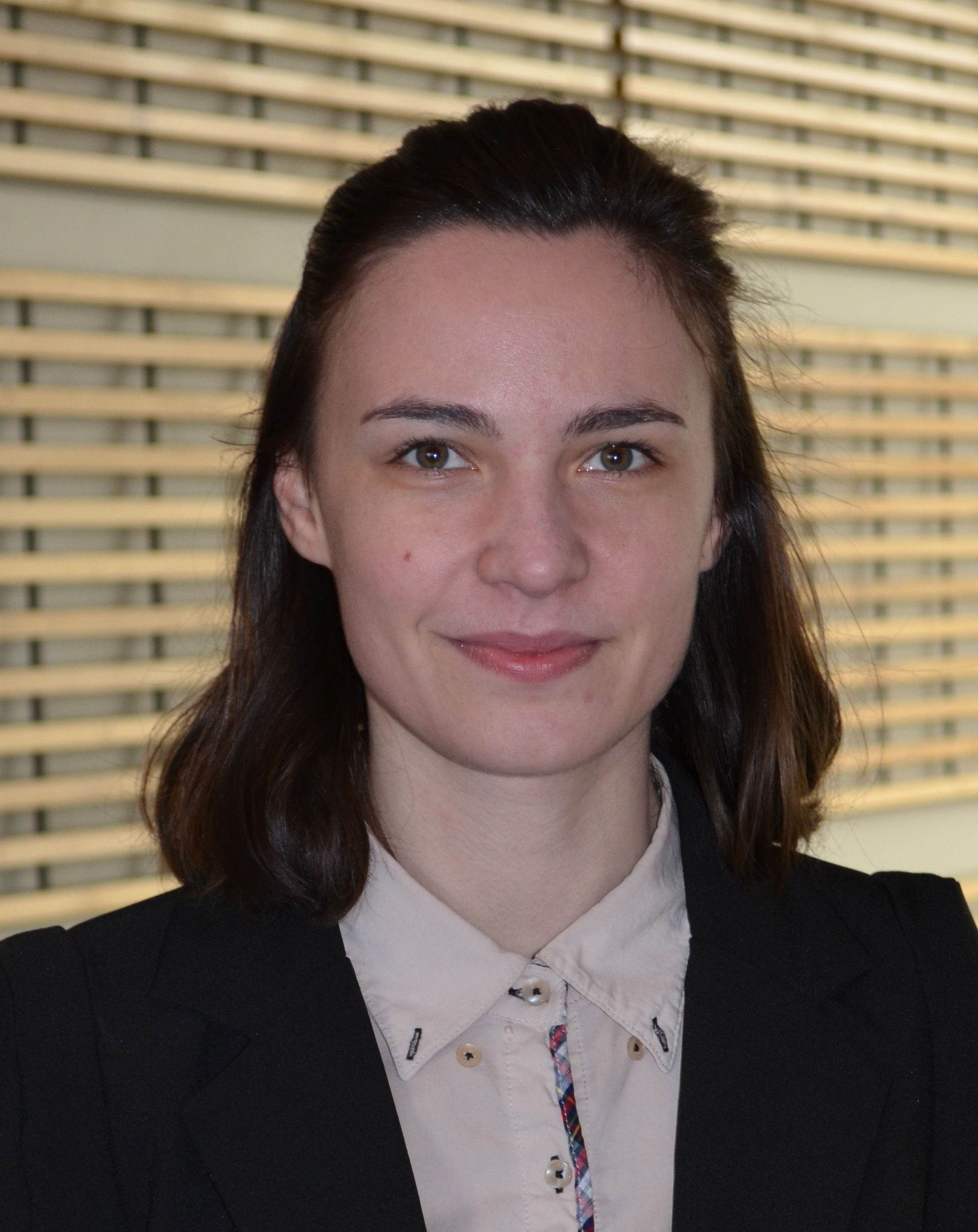}}]
 	{Mihaela Dimovska} 
 	holds a master's degree in Computer Science from the University of Tennessee, Knoxville and a bachelor's degree in Computer Science and Mathematics from the American University in Bulgaria. 
    She was a research assistant at the University of Tennessee, Knoxville from 2016 till 2018. In 2019 was a research fellow in the Advanced Short Term Research Opportunity (ASTRO) at Oak Ridge National Laboratory. Currently, she is a research assistant and a Ph.D. candidate at the University of Minnesota. 
    Her research interests include system identification, graphical models, machine learning, and stochastic modeling. 
\end{IEEEbiography}
\vspace{-0.1in}
 \begin{IEEEbiography}[{\includegraphics[width=1in,height=1.25in,keepaspectratio]{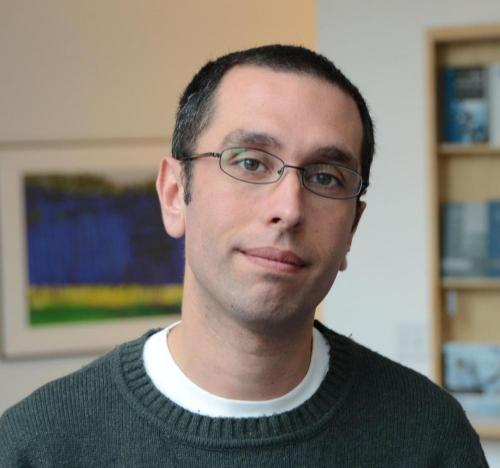}}]
 	{Donatello Materassi} holds a Laurea in ``Ingegneria Informatica'' and a ``Dottorato di Ricerca'' in Electrical Engineering/Nonlinear Dynamics and Complex Systems from Universit\`a degli Studi di Firenze, Italy.
 	He has been a post-doctoral researcher at University of Minnesota (Twin Cities) from 2008 till 2011 and at the Massachusetts Institute of Technology (MIT) from 2011 till 2014. While at MIT he has concurrently been a lecturer at Harvard University.
 	Since 2014 he has been an assistant professor at University of Tennessee in Knoxville and since 2019 at University of Minnesota.
 	He received the NSF CAREER award in 2015.
 	His research interests include nonlinear dynamics, system identification and classical control theory with single molecule force spectroscopy, biophysics, statistical mechanics and quantitative finance.
\end{IEEEbiography}
\end{document}